\documentclass{aastex}
\usepackage{psfig}
\usepackage{emulateapj5}
\usepackage{apjfonts}

\def\nh21{n_{\rm H,21}}
\def\flux{F_{-12}}
\begin{document}
\slugcomment{Submitted to ApJ Letters: July 17; revised August 5, 2001}
\title{ 
The Compact Central Object in the RX~J0852.0--4622 Supernova Remnant}
\author{ 
George G.\ Pavlov, Divas Sanwal, B\"ulent K{\i}z{\i}ltan, 
and Gordon P. Garmire
}
\affil{
The Pennsylvania State University,
525 Davey Lab, University Park, PA 16802
}
\email{pavlov@astro.psu.edu; divas@astro.psu.edu}
\begin{abstract}
The central region of the recently discovered supernova remnant  
RX~J0852.0--4622
was observed with the ACIS detector aboard the {\sl Chandra} X-ray
Observatory.  We found only one relatively bright source,
about $4'$ north of the SNR center, with a flux of 
$\sim 2\times 10^{-12}$ erg~s$^{-1}$~cm$^{-2}$ in the 0.5--10 keV band.
The position of this point-like source, CXOU J085201.4--461753, rules out
its association with the two bright stars in the field, HD 76060 and
Wray 16-30. Observations of the field with the CTIO 0.9-m telescope show
a star ($R\approx 17$, $B\approx 19$)
at about 2\farcs4
from the nominal X-ray position. 
We consider association of this star with the X-ray source unlikely
and estimate a limiting magnitude of the optical counterpart as
$B \ge 22.5$ and $R \ge 21.0$.  Based on the X-ray-to-optical flux
ratio, we argue that the X-ray source is 
likely the compact
remnant of the supernova explosion that created
the RX~J0852.0--4622 SNR. The observed X-ray spectrum of the source
is softer than  spectra of magnetospheric radiation of rotation-powered
pulsars, but it is harder than spectra of cooling neutron stars emitting
thermal radiation from the entire surface,
similar to the central compact source of the Cas A SNR.  
We suggest that CXOU J085201.4--461753 belongs to the
growing family of radio-quiet compact central sources, presumably
neutron stars, recently discovered in a number of SNRs. 
\end{abstract}
\section{Introduction}
The shell-like supernova remnant (SNR) RX~J0852.0--4622 (G266.2--1.2)
at the south-east corner of the Vela SNR was 
discovered 
by Aschenbach (1998) using data from the {\sl ROSAT} All-Sky Survey (RASS).
High 
temperatures, $\ga 3\times 10^7$ K, obtained by Aschenbach under the
assumption that the SNR spectrum is thermal,
suggested that this is a young SNR
($\tau\la 1,500$ yr). Possible detection of the 1.156 MeV $\gamma$-ray
line of the  radioactive isotope $^{44}$Ti (half-life $\sim 90$ yr) with
{\sl CGRO} COMPTEL (Iyudin et al.\ 1998) may imply an even younger age,
$\sim 680$ yr, at a distance of $\sim 200$ pc,  estimated from the SNR
age and diameter, $\sim2\arcdeg$. Aschenbach, Iyudin, \& Sch\"onfelder
(1999) estimate upper limits of 1,100 yr for the age and 500 pc for the
distance.  Thus, RX~J0852.0--4622 could be the remnant of the nearest 
supernova to have occurred in the recent human history. 
However, further observations with {\sl ASCA} 
(Tsunemi et al.\ 2000; Slane et al.\ 2001) have shown
that the X-ray spectra of the SNR shell are nonthermal (i.e., the
SNR is not necessarily very hot). Moreover, power-law fits of the spectra
 suggest
that the hydrogen column density toward RX~J0852.0--4622 is an order of
magnitude higher than that for the Vela SNR, implying that a plausible
distance to the remnant is 1--2 kpc, so that an age of a few thousand years,
inconsistent with the reported 
$^{44}$Ti line flux,
is more likely. 

Aschenbach et al.\ (1999) suggest that RX~J0852.0--4622 was created by
a core-collapse supernova which left 
a neutron star (NS)
or a black hole (BH).  Aschenbach (1998) found a point-like source
(A1 hereafter) in the RASS data with a flux of $F_{-12}\sim 8$
(where $F_{-12}\equiv F/[10^{-12}~{\rm erg~cm}^{-2}~{\rm s}^{-1}]$),
off-set by $2'$--$5'$ to the 
north from the SNR center. A second RASS source (A2),
even closer to the center, with a
countrate of 0.12 s$^{-1}$,  was reported by Aschenbach
et al.\ (1999). If A1 or A2 is a NS, the low detected fluxes
imply a very low surface temperature (e.g., $3\times 10^5$ K for
$n_{\rm H}=10^{20}$ cm$^{-2}$, $d=400$ pc, and NS radius $R=10$ km) or a large
distance or a small emitting area.
  Slane et al.\ (2001) observed an unresolved central source (S)
surrounded by diffuse emission; its position is marginally consistent with
that of A1. The blackbody (BB) fit of the S spectrum gives $kT=0.47\pm 0.04$
keV,
$\nh21\equiv n_{\rm H}/(10^{21}~{\rm cm}^{-2}) < 0.8$, and unabsorbed flux
$\flux^{\rm (un)} = 2.1$ in the 0.5--10 keV band. The power-law (PL) fit
gives a photon index $\gamma=3.2\pm 0.5$, 
$\nh21=1.1$--7.5, and $\flux^{\rm (un)} =6.6$.
Mereghetti (2001) observed the field with {\sl BeppoSAX} and detected a source
(M1) whose position is consistent with those of A1 and S.
The spectral parameters and the flux values [$kT=0.5\pm 0.1$ keV, 
$F_{-12}^{\rm (un)}(2$--$10~{\rm keV})=0.56$  for the BB fit;
$\gamma=3.6\pm 0.6$, $F_{-12}^{\rm (un)}(2$--$10~{\rm keV})=0.66$ 
for the PL fit] are consistent with those obtained by Slane et al.\ (2001). 
Mereghetti (2001) also reports detection of another source (M2), about $3'$
north of M1, which is seen at a $7\sigma$ level at $E>5$ keV. The flux of M2 is
estimated as $\flux\sim 0.3$--0.4 in the 2--10 keV band. Because of its harder
X-ray spectrum and higher X-ray-to-optical flux ratio, Mereghetti considers
M2 as a more likely candidate for the SNR central source.

Based on the close positions of A1, S and M1 and
similar spectra of S and M1, it is reasonable to assume that it is
actually the same source. Its nature has remained elusive because there
are several stars within the X-ray error circles, including two bright
early-type stars, HD 76060 ($V=7.88$) and Wray 16-30 ($V=13.8$),
which could provide the observed X-ray flux.  Alternatively, this source
could be a background AGN or a compact remnant of the SN explosion, although it is not easy
to reconcile its properties with those of ``canonical'' young NSs. 
The nature of two
other sources, A2 and M2, is even more uncertain. 
To understand which (if any) of the three
sources is a viable candidate for the compact remnant and infer its
properties, we proposed observations of this field with the {\sl Chandra}
X-ray Observatory ({\sl Chandra}).  In this Letter we present the results of
a 3 ks snapshot  observation with the {\sl Chandra} Advanced CCD Imaging
Spectrometer (ACIS; Garmire et al.\ 1992; Bautz et al.\ 1998) and report 
optical observations of this field.
\section{ACIS Observation}
The {\sl Chandra} ACIS observation was taken on 2000 October 26 (2,994~s
effective exposure).  The central part ($\approx 28'\times 17'$) of the
SNR was imaged on four chips of the imaging array
ACIS-I and two chips of the spectroscopic array ACIS-S.  The data were
acquired in Timed Exposure mode, with the default frame readout time
of 3.24 s.

Figure~1 shows a $12'\times 12'$ 
region of the ACIS-I image
with the only point-like source (marked X) clearly detected in
this observation, together with the error circles for the previously
claimed detections of point sources.
The source CXOU~J085201.4--461753 (we will call it Source X for brevity) 
is centered at 
RA(J2000)$=8^{\rm h}52^{\rm m}01\fs38$,
Dec(J2000)$=-46^\circ 17' 53\farcs34$.  For the processing software
version used, the rms 
error of celestial location\footnote{
{\tt http://asc.harvard.edu/mta/ASPECT/cel\_loc/cel\_loc.html}}
is smaller than $0\farcs6$, although in some cases the offset can be
as large as $2''$.  Source~X lies within the error circles of 
A1, S and M1, so that it is likely the same source.  
It is imaged on chip I2, about
$2\farcm5$ off-axis.  Although it is embedded in a strip of enhanced
background, no sign of a small-scale structure is seen around the source.

\vskip 5truept
\hbox{
\hskip 5pt
\psfig{figure=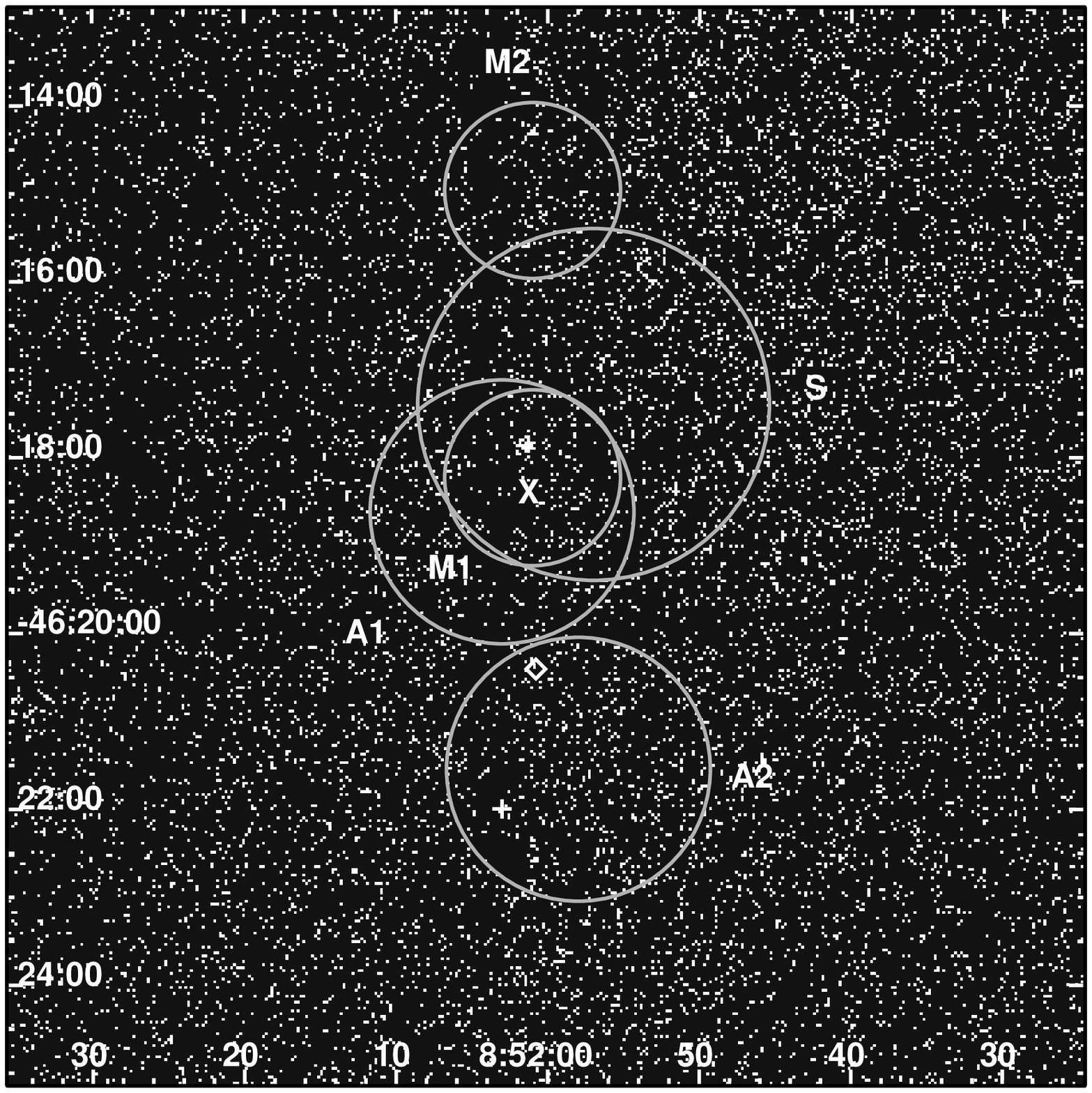,width=8.5cm}
}
\figcaption{
X-ray image, $12'\times 12'$,
of the RX~J0852--4622 central region.
The symbols X, +, and $\Large \diamond$ mark the source
CXOU J085201.4--461753 (Source X), the SNR center, and
the ACIS-I aimpoint, respectively.
The five error circles for the previously claimed point source
detections with {\sl ROSAT} (A1 and A2), {\sl ASCA} (S), and
{\sl BeppoSAX} (M1 and M2) are shown (see text for details).
}
\vskip 10truept
 
We 
found 448 photon events (standard grades 02346)
within a $2\farcs5$ radius circle (which
includes about 95\% of point source counts).  We estimate that such an
aperture contains, on average, 0.56 background counts.
Thus, the observed source countrate (in the 2\farcs5 aperture) is 
$0.150\pm0.007$ s$^{-1}$.

Measuring the spectrum and the flux of the source is complicated due to two
reasons. First, 
the charge transfer inefficiency (CTI)
distorts the spectrum and reduces the energy resolution,
particularly for sources imaged on a front-illuminated
chip far from the frame store, like our source.
To correct for this distortion,
we applied the CTI corrector (Townsley et al.\ 2000) to the Level 1 data.
The CTI-corrected spectrum is somewhat harder than the uncorrected one,
although the difference in spectral parameters is of the order of
statistical uncertainties.

More severe distortion of the data is caused by CCD pile-up\footnote{
see The {\sl Chandra} Proposers' Observatory Guide (POG), v.3.0, pp.103-108;
{\tt http://asc.harvard.edu/udocs/docs/docs.html}},
which 
results in a reduced count rate
in the
image core and hardening of the spectrum.  The observed countrate of
our source, 0.485 counts/frame, is high enough for the pile-up effects
to be quite substantial. Indeed, using the spectral parameters inferred
from the {\sl ASCA} and {\sl BeppoSAX} data, we estimated, with the aid
of PIMMS\footnote{\tt http://heasarc.gsfc.nasa.gov/Tools/w3pimms.html},
that the ACIS-I countrate of 0.15 s$^{-1}$ after pile-up corresponds
to 0.29--0.30 s$^{-1}$ (0.94--0.97 counts/frame) before pile-up.
Therefore, the spectral parameters obtained from  the ACIS spectra
neglecting pile-up [$\nh21 = 3\pm 1$, $kT=0.53\pm 0.06$ keV, 
$R=(0.1$--$0.2)d_{\rm kpc}$ km, $\flux(0.5$--$10~{\rm keV})
=1.0$ for the BB fit; $\gamma=2.7\pm 0.5$,
$\nh21=8\pm 2$,  $\flux(0.5$--$10~{\rm keV})=1.4$ for
the PL fit] are inaccurate.

We attempted to evaluate the spectral parameters 
using the codes {\tt XPSF} 
(Chartas 2001, in preparation) and {\tt ISIS} (Davis 2001)
to correct for the pile-up effects.  Unfortunately,
these codes are not very efficient when the number of source counts
is small, so that we obtained only crude estimates:
$\nh21=8$--12,        
$\gamma = 3$--5,       
for the PL fit;
$\nh21=2$--6,         
$kT=0.3$--0.5 keV,    
for the BB fit; the observed flux $\flux(0.2$--$10~{\rm keV}) =1$--3
in both cases.  The spectral parameters are compatible with those obtained by
Slane et al.\ (2000) and Mereghetti (2001). 
The $n_{\rm H}$ value obtained in the PL fit 
is close to the total galactic HI column density
in this direction, $\approx 1\times 10^{22}$ cm$^{-2}$ (Dickey \& Lockman 1990;
estimated with the aid of the  
{\tt w3nH}
tool\footnote{\tt http://heasarc.gsfc.nasa.gov/cgi-bin/Tools/w3nh/w3nh.pl}).
The BB fit yields a lower $n_{\rm H}$, but it is still 
considerably greater than the highest value, $\nh21=0.6$, found by Lu \& 
Aschenbach (2000) for the southern part 
of the Vela SNR.
Since BB fits give lower $n_{\rm H}$ values than any other simple
spectral model, we conclude that the observed source is well beyond
the Vela SNR
(the distance to
Vela is $\approx 250$ pc --- Cha, Sembach, \& Danks 1999),
but it should be closer than the Vela Molecular Ridge at a distance of $\sim 1$--2 kpc
(see Slane et al.\ 2000 for a more detailed discussion).

To search for other sources in the image, we have used the {\tt wavdetect}
tool\footnote{{\tt http://asc.harvard.edu/ciao/threads/detect.thread.html\#RunWV}}.
We find only two sources in the full-band image, the Source~X
and another source with a statistical significance of 2.4$\sigma$.
Since Mereghetti (2001) found the source M2 at higher energies only,
we divided the data into two energy ranges $E < 2$ keV and $E > 2$ keV,
and ran {\tt wavdetect} on the two 
images separately. We find that only the
Source~X shows up in both 
images.  In the low energy
image, there is an additional source detected  at
RA(J2000)$=8^{\rm h}52^{\rm m}25\fs38$, Dec(J2000)$=-46^\circ 15' 55\farcs4$,
with total $7$ counts and a statistical significance of 3.1$\sigma$. In the
high energy image, no other statistically significant source is found. No
sources are detected by {\tt wavdetect} at positions compatible with 
sources A2 and M2.

To put an upper limit to the flux at the positions corresponding to the
sources A2 and M2, we examined the brightest point-like background
enhancements within the error circles corresponding to the two sources.  
We found upper limits of 4 and 5 source counts in a $2\farcs5$ radius
extraction circle for A2 and M2, respectively.
These upper limits are too low to be consistent with the reported
countrates for A2 and M2. For instance, using the parameters
$\gamma=1$--3, $\nh21=4$, $\flux(2$--10~keV)=0.3--0.4, 
assumed by Mereghetti (2001) for the M2 source, we estimate
that ACIS-I would detect 50--300 counts in a 3 ks exposure.
Similarly, the PSPC countrate of 0.12 s$^{-1}$, reported by Aschenbach
for A2, is incompatible with the ACIS-I upper limit of 0.0013 s$^{-1}$
at any reasonable spectral parameters. 
Thus, we have to conclude that A2 and M2 are either highly variable
sources or
misdetections.
Therefore, we believe there is only one viable candidate
--Source X-- for the compact remnant.
\section{Search for Optical Counterpart}
We have surveyed optical catalogs to search for a possible optical
counterpart of Source X.
The two brightest
optical stars in the field, HD 76060 and Wray 16-30, 
are immediately excluded as counterpart candidates because they
are $34''$ and $24''$ away from the X-ray source, respectively.
No counterparts were found in the USNO A-2 catalog ---
the nearest object is $12''$ away.  The Digitized Sky Survey (DSS)
plates, obtained with the UK Schmidt Telescope (UKST),
show one possible candidate (we will call it Star Z) 
whose position, determined with an accuracy of about $1''$,
differs from the Source X position  by $2\farcs3$.
We found a record for this source in the ROE/NRL COSMOS UKST 
Southern Sky Object Catalog (Yentis et al.\ 1992\footnote{The online
interactive  catalog is available at
{\tt http://xip.nrl.navy.mil/www\_rsearch/RS\_form.html}}).
According to this catalog, Star Z has a B magnitude of 18.7,
and its position differs by $3\farcs 3$ from the Source X position.

The field around Source X 
was observed, at our request, by
J.~Allyn Smith 
with the CTIO 0.9-m telescope . Three 10-min and three 5-min
exposures with the Sloan DSS (SDSS)
 g, r and i filters, taken on 2001 March 27 with the
CCD detector TEK2K\_3 (plate scale $0\farcs 4$ per pixel), allow us to
calculate the position of Star Z more accurately and estimate its
colors. We obtained a plate solution based on USNO A-2 positions of
22 field stars and found the coordinates of Star Z:
RA(J$2000)=8^{\rm h}52^{\rm m}01\fs17$,
Dec(J$2000)=-46^\circ 17'52\farcs3$,
with an accuracy of $0\farcs2$. 
Taking $0\farcs6$ as a nominal uncertainty of the 
{\sl Chandra} source position,
the optical to X-ray offset of
2\farcs4
(2\farcs2 west, 1\farcs 0 north --- see Fig.~2) 
is significant at about $3.8 \sigma$ level.
However, we cannot exclude the possibility that the actual error of the
X-ray position can be as large as the offset observed.

Since no standards were observed during the CTIO observations, we
cannot directly measure the true magnitudes of Star Z.  However,
plots of empirical color-magnitude dependences
for the observed field show that Star Z
is well within the strips formed by field stars, i.e.,
most likely it is a main-sequence star.
To estimate its true magnitudes and colors,  we found an object
in the CTIO images with approximately the same apparent g, r, i
magnitudes  as Star Z (the differences do not exceed 0.12 mag) 
and identified this object in the USNO A-2 catalog, which gives
$B = 18.9$, $R = 16.9$. (The difference of 0.2 mag between the B magnitudes
of this object and Star Z is less than the typical photometric
error of USNO A-2 of about 0.25 mag).
We also measured the g, r and i magnitudes of 45 field stars for
which the USNO B and R magnitudes are known.
Using the transformation from the SDSS system to the Johnson--Morgan--Cousins
system (Fukugita~et~al. 1996), we transformed the measured g, r and i
magnitudes to corresponding B and R magnitudes
and computed a 
linear regression fit 
for a transformation from
the converted B and R magnitudes to the USNO B and R magnitudes.
Using this fit, we found
$B = 18.8 \pm 0.4$ and $R=16.8 \pm 0.5$ for Star Z, compatible with
those estimated above.

\vskip 5truept
\hbox{
\hskip 5pt
\psfig{figure=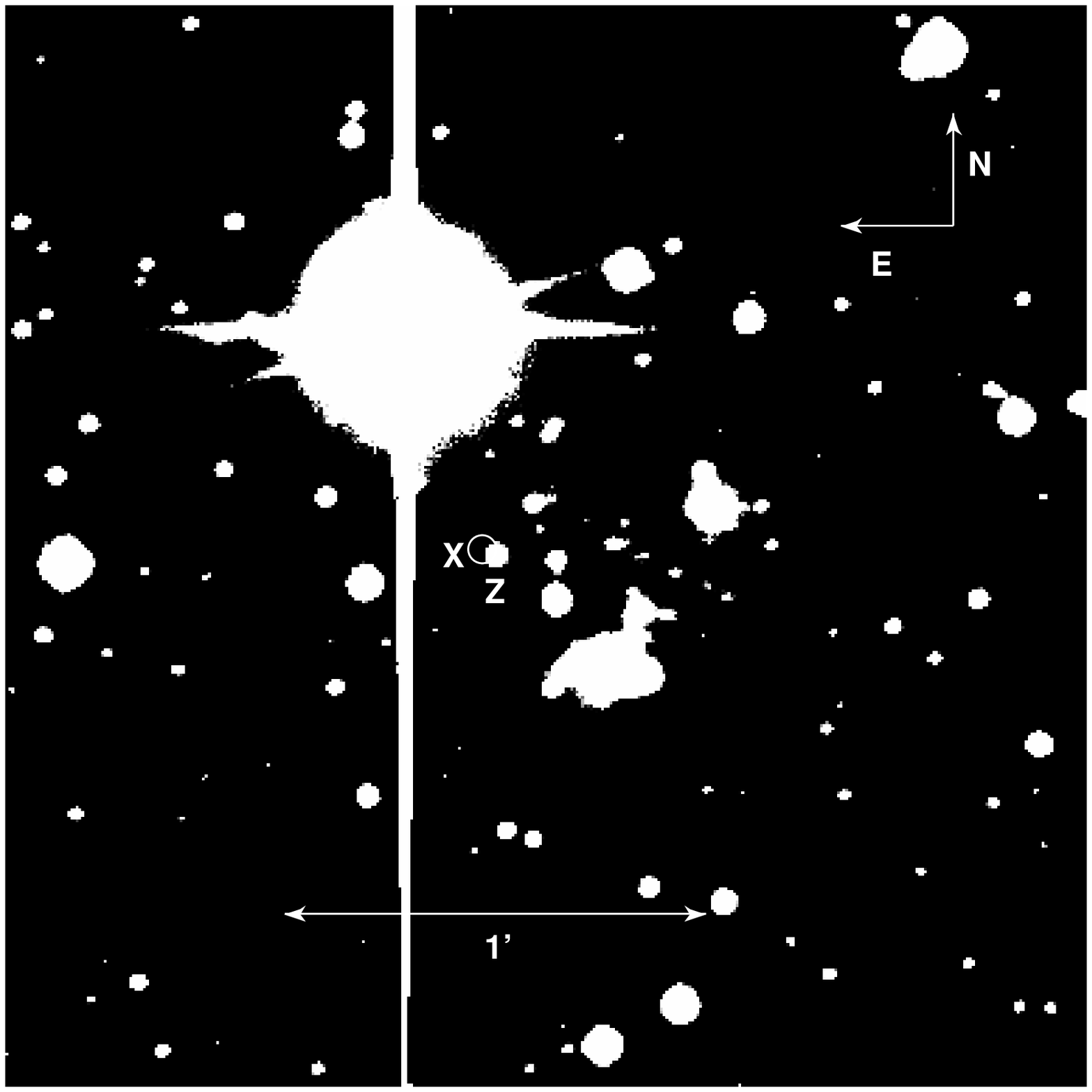,width=8.5cm}
}
\figcaption{
The
r band
image, about $2\farcm5 \times 2\farcm5$,
of the central region of the SNR RX~J0852--4622
(10 min exposure).
The {\sl Chandra} position of Source X is shown by a circle of $2''$ radius
near the center of the image.  The nearest star, 2\farcs4 south-west of
Source X is Star Z.  Two brightest stars in the image are HD 76060 and
Wray 16-30, north-east and south-west of Source X, respectively.
}
\vskip 10truept

Adopting these magnitudes for Star Z, we could
estimate its spectral type and distance if the interstellar extinction
were known. The extinction could be estimated from the $n_{\rm H}$ value
(e.g., $A_B\approx 0.7\,\nh21$, $A_R\approx 0.4\,\nh21$ --- 
Savage \& Mathis 1979),  but
this value remains
highly uncertain.  For instance, if we assume $\nh21=3$ (compatible
with the BB fit of the X-ray spectrum), then the dereddened magnitudes
are $B_0=11.8$, $R_0=15.7$, and Star Z could be
a G0V star at $d=2.3$ kpc. At $\nh21=1$ (a minimum value compatible
with the X-ray spectrum), it could be a K2V star at $d=1.5$ kpc.  
If $\nh21$ is as high as 10 (as obtained from the PL fit), then
$B_0-R_0=-1.1$, i.e., the dereddened spectrum is so blue that it cannot
be 
the radiation from an ordinary star. 
Moreover, since $\nh21\ga 10$
for the total galactic hydrogen column density, 
this estimate of the dereddened color
would be applicable if Star Z were an extragalactic source (e.g. an AGN).
However, even bluest AGNs have $B_0-R_0 >-0.2$
(D.~Schneider, private communication), which means that Star Z is
not an AGN, too, i.e., such an assumption on the $n_{\rm H}$ value
is unrealistic.

If Star Z is not the optical counterpart, then limiting
magnitudes corresponding to the X-ray source position can be estimated
as $B \ge 22.5$ and $R \ge 21.0$ from the CTIO 0.9 m images.
\section{Discussion}
The  nature of Source X
can be constrained from the ratio of  its X-ray
and optical fluxes, $F_x/F_{\rm opt}$.  Although we do not consider Star
Z as a plausible optical counterpart, based on its colors and
positional offset, we cannot completely rule out the association
and will use its optical flux to obtain a lower limit on $F_x/F_{\rm opt}$.
With $\flux(0.5$--$10~{\rm keV})>1$, $B=18.8$, $R=16.8$, we obtain
$F_x/F_B>5$, $F_x/F_R>1.3$
 for the observed fluxes (i.e., uncorrected
for interstellar extinction). The ratios for unabsorbed fluxes strongly
depend on the poorly known hydrogen column density and
the model assumed for the X-ray spectrum. For example,
assuming a BB model with $kT=0.4$ keV, $\nh21=3$, we obtain 
$F_x^{\rm (un)}/F_B^{\rm (un)}>1.1$, 
$F_x^{\rm (un)}/F_R^{\rm (un)}>0.6$ if the unabsorbed
X-ray flux is estimated for the 0.5--10 keV band.  The estimated lower
limits are high enough to conclude that 
Source X is not
a main-sequence star, which means that if even Star Z is a main-sequence
optical counterpart, X-rays emerge from a different object
(e.g., from a compact companion of a binary system).
The estimated
flux ratio does not rule out the possibility that the 
putative optical-X-ray
source is a magnetic cataclysmic variable (mCV) --- $F_x/F_{\rm opt}$
can be as high as $\sim 300$ for such objects (see, e.g., Warner 1995 for
references).  However, the observed X-ray spectrum is too hard to be 
a soft component ($kT\sim 0.02$--0.04 keV) of an mCV spectrum, 
and it is too soft to be a
hard component (dominant, typically, at $E\ga 0.5$ keV) 
which usually fits
a thermal bremsstrahlung model with $kT_{\rm br}\approx 20$--40
keV (versus  $kT_{\rm br}=1$--2 keV for Source X).  In addition, if
it were an mCV, its radiation should be strongly variable,  whereas
the observed {\sl ASCA}, {\sl BeppoSAX}, and {\sl Chandra} fluxes do not
show appreciable differences.

If Star Z is just a field star, not related to Source X, then 
$F_x/F_B>150$, $F_x/F_R>60$,
which excludes any star, except for a NS or a BH, as a source of the observed
X-ray radiation. One could assume that
Source X is an AGN, heavily absorbed by the interstellar material in
the galactic plane. In this case, the unabsorbed soft-X-ray flux
is so large [e.g., $F^{\rm (un)}(0.1$--$2.4~{\rm keV})=
4.3\times 10^{-10}$ erg~cm$^{-2}$~s$^{-1}$ for $\flux(0.5$--$10~{\rm keV})=2$,
$\nh21=10$, $\gamma=4$] that such an AGN would be seen as an extremely
bright source if it were at a high galactic latitude (e.g., the {\sl ROSAT}
PSPC countrate would be about 15 s$^{-1}$ for $\nh21=0.2$ at the flux and
photon index assumed above). The probability to find such a bright AGN,
which, in addition, has a spectrum much steeper,
and $F_x^{\rm (un)}/F_{\rm opt}^{\rm (un)}$ much larger than a usual AGN,
within an area of a few square arcminutes is practically negligible.
Thus, we conclude that Source X is not a field AGN, and thus 
very likely it
is the compact remnant of the supernova explosion.

Given the steep slope of the X-ray spectrum, low luminosity
[$L_x(0.5$--$10~{\rm keV}) \sim (10^{32}$--$10^{34})d^2_{\rm kpc}$ erg~s$^{-1}$
for the PL model],
and the lack of a pulsar-wind nebula around Source X, we do not expect
it to be a young active (radio- and/or $\gamma$-ray) pulsar. On the
other hand, its BB temperature, $kT\sim 0.4$ keV, is too high for a cooling
NS of a reasonable age, and the emitting radius, $R\sim 0.4 d_{\rm kpc}$
km, is too small to explain the radiation as emitted from the whole
surface of an isolated NS.   Interestingly, the spectral properties of
Source~X 
resemble those of the compact central object in the
very young Cas A SNR --- e.g., the BB fit of the Cas A source
spectrum gives  
a slightly higher temperature,
$\sim 0.6$ keV, with a comparable emitting area and luminosity
(Pavlov et al.\ 2000). The nature of the Cas A central source remains
elusive. Various hypotheses (an isolated NS with hot spots,
an anomalous X-ray pulsar, a compact
object slowly accreting from a fossil disk or a dwarf binary
companion),  which also pertain to Source~X
and some other radio-silent NSs in SNRs, have been discussed in literature
(Murray et al.\ 2001, and references therein).

To conclude, the X-ray and optical observations show that 
CXOU~J085201.4--461753, the only bright X-ray object close to the center
of the RX~J0852--4622 SNR, is most likely a compact remnant of the supernova
explosion.
Its properties are different from those of a classical
rotation-powered pulsar or a cooling NS with a uniformly heated surface.
Likely, it is a member of the newly emerging class of radio-
(and $\gamma$-ray) quiet young isolated NSs in SNRs.  To elucidate the
nature of this source, additional X-ray observations are required to get
a more accurate spectrum and to search for periodic and aperiodic variability.

\acknowledgements{
We are grateful to J.~Allyn Smith for sharing observation time at CTIO
and to Robin Ciardullo for arranging the CTIO observation. 
We thank George Chartas and John Davis for providing their codes
to correct for CCD pile-up, Leisa Townsley for the useful advice
on CTI correction, and Bernd Aschenbach, the referee, for useful
remarks.
This research was supported by NASA grants NAS8-38252 and NAG5-10865.
}

\end{document}